\documentstyle[11pt,newpasp,epsf]{article}

\begin{document}

\title{Radiative transfer in 3D model stellar atmospheres}

\author{Martin Asplund}
\affil{Research School of Astronomy and Astrophysics, 
Mt Stromlo Observatory, Cotter Road, Weston, ACT 2611, Australia
(martin@mso.anu.edu.au)
}

\author{Remo Collet}
\affil{Institutionen f\"or Astronomi och Rymdfysik,
Box 515, SE-751 20 Uppsala, Sweden (remo@astro.uu.se)}
\begin{abstract}

Recently 3D hydrodynamical simulations of stellar surface convection
have become feasible thanks to advances in computer technology and
efficient numerical algorithms. Available observational diagnostics indicate
that these models are highly realistic in describing the topology of
stellar granulation and for spectral line formation purposes.
The traditional free parameters (mixing length parameters,
micro- and macroturbulence) always inherent in standard 1D analyses
have thus finally become obsolete.
These 3D models can therefore both shed light on the elusive nature of
stellar convection as well as be employed in element abundance analyses.
In the present contribution we will describe some aspects of the models and
possible applications of them in terms of radiative transfer.

\end{abstract}

\keywords{Stellar abundances, stellar convection, radiative transfer}

\section{Introduction}

Reality is (at least) 3-dimensional. This obvious statement is equally
applicable to stellar atmospheres, in particular those of late-type stars
where the surface convection zone reaches the region responsible for the emitted
stellar radiation. Even a casual glimpse of the surface of our own Sun
immediately reveals the existence of photospheric inhomogeneities in the form
of granulation, in addition to magnetic phenomena like sunspots and flux tubes
which probably at some level also are related to the convective motions. Granulation is
the direct observational manifestation of convection: concentrated, rapid
downdrafts of cool material (intergranular lanes)
in the midst of broad, slow upwellings of warm gas (granules).
Furthermore, the solar atmosphere is distinctly time-dependent: the individual granules are
continuously evolving with
typical life-times of a few minutes and waves excited by the convective motions
are regularly propagating into the upper layers of the photosphere and the chromosphere.
Similar features are expected on other
late-type stars and may in fact often be more severe than in the Sun.
In view of the dynamic nature of stellar atmospheres, one has to worry whether
the use of time-independent hydrostatic 1D model atmospheres as regularly done in
for example stellar abundance analyses could lead to severe systematic errors.
Here we will describe efforts currently being made in developing more physically motivated
models of stellar atmospheres and the line formation processes, which
should hopefully place the findings on firmer footing.

With the advent of supercomputers it is now feasible to perform highly realistic
time-dependent 3D hydrodynamical simulations of stellar surface convection
(e.g. Nordlund \& Dravins 1990; Stein \& Nordlund 1998; Asplund et al. 1999, 2000a,b,c;
Asplund \& Garc\'{\i}a P{\'e}rez 2001).
Without invoking any of the free parameters traditionally employed in 1D analyses
the 3D simulations successfully reproduce a large number of observational constraints
such as the granulation topology and detailed line shapes and asymmetries.
In this context the energy exchange between the gas and the radiation field
is absolutely crucial by largely determining the atmospheric
structure and ultimately driving the convection. In the present contribution,
special emphasis will be placed on how realistic radiative transfer is achieved
both in constructing the 3D convection simulations and in predicting the emergent
spectra since different approaches, assumptions and approximations are necessary in
the different cases.
It is important to emphasize that although the present 3D models
are not perfect, the main limitations are of a computational nature, which will be
naturally addressed in the future when more powerful computers become available,
and due to a lack of manpower.
In contrast, 1D models are flawed in a physical sense, since the intrinsic
restriction to 1D can not adequately describe all aspects of what is inherently a
3D phenomenon.
That is of course not to say that 1D models do not have their merits but it
is important to investigate their potential shortcomings, which can only be
achieved by comparisons with more sophisticated modelling.

\section{3D Hydrodynamical Simulations of Stellar Surface Convection}

\subsection{Ingredients}

The 3D hydrodynamical simulations of stellar surface convection
described herein have been
computed with a 3D, time-dependent, compressible, explicit,
radiative-hydrodynamics code
(Stein \& Nordlund 1998). The hydrodynamical
equations for conservation of mass, momentum
and energy are solved on a non-staggered Eulerian
mesh with gridsizes of $\approx 100^3$. The physical
dimensions of the grids are sufficiently large
to cover many ($>10$) granules simultaneously
in the horizontal direction and many ($>10$) pressure
scale-heights in the vertical.
The simulations are thus only covering a small but
representative fraction of
the whole stellar surface and convection zone, which
gives statistically significant results for long time sequences.
In terms of
continuum optical depth the simulations extend
at least up to log\,$\tau_{\rm Ross} \approx -5$ which for
most purposes are sufficient to avoid
numerical artifacts of the open upper boundary on
spectral line formation which could otherwise skew the results.
The lower boundary is located at large depths to
ensure that the inflowing gas is isentropic and featureless,
while periodic horizontal boundary conditions are employed.
The temporal evolution of the simulations cover
several convective turn-over time-scales
to allow thermal relaxation to be established.
The simulations include a hyper-viscosity for stabilization
purposes whose parameters are fixed from standard
hydrodynamical test cases.
No magnetic fields have been included although the code is capable of
simulating magneto-convection.
The simulations described here thus only apply to magnetically
inactive stars such as the Sun.

Since the intention is to make the simulations as realistic
as possible, state-of-the-art input physics is used.
The adopted equation-of-state comes from Mihalas et al. (1988),
which accounts for ionization, excitation and dissociation of
the most important atoms and molecules. The Uppsala opacity
package (Gustafsson et al. 1975 with subsequent updates) provide
the continuous opacity sources and the extensive
lists of Kurucz (1998, private communication) the line opacities.
The input parameters characterizing the models are the surface
gravity (log\,$g$), the overall metallicity ([Fe/H]) and the entropy
at the lower boundary. The effective temperature ($T_{\rm eff}$)
is a consequence of the granulation properties, which
depend on the entropy structure.
To construct a model with a prespecified $T_{\rm eff}$ is therefore
a rather time-consuming task involving making several simulations until
exactly the right entropy at the bottom has been found which produces
the targeted $T_{\rm eff}$.
An alternative approach is to settle for a $T_{\rm eff}$ in the vicinity
of the desired one and carry out a differential analysis with 1D models
with exactly the same stellar parameters. This procedure is justified since
3D effects will not change drastically with minor adjustments of the
parameters and has been adopted in most cases by us.
The reader is referred to Stein \& Nordlund (1998),
Asplund et al. (1999, 2000a,b) and
Asplund \& Garc\'{\i}a P{\'e}rez (2001)
for further details of the 3D convection simulations employed here.

\subsection{Radiative Transfer}

An accurate representation of the energy balance between the gas
and the radiation field is crucial in order to achieve a realistic
atmospheric structure. We therefore compute detailed 3D radiative
transfer at each time-step for a selected number of inclined rays
(typically around 10) using a Feautrier-like (1964) long characteristic formal solution.
To improve the numerical accuracy, the radiative transfer is solved for
a finer depth grid and the results subsequently interpolated back to
the original hydrodynamical grid.
For computational reasons, the assumption of local thermodynamic
equilibrium (LTE) without continuous scattering terms
($S_\lambda = B_\lambda$) is made throughout.
The effects of line-blanketing is accounted for through opacity binning
(Nordlund 1982) using normally four opacity bins. The original
2748 wavelength points are sorted into groups representing continuum,
weak lines, intermediate strong lines and strong lines. For each
bin pseudo-Planck functions are computed and stored in a lookup-table
together with the opacities and equation-of-state. Even with the much simplifying
assumptions of LTE and only four opacity bins, it is noteworthy that the
most time-consuming task in the simulations is the solution of the
radiative transfer while the time-integration of the hydrodynamical
equations only require a fraction of the time for the highest
resolution simulations.
We are currently designing methods to switch to a selected opacity
sampling technique with $50-100$ wavelength points instead of opacity binning
(Trampedach et al., these proceedings).

\begin{figure}[t!]
\label{Tv_slice}
\plotone{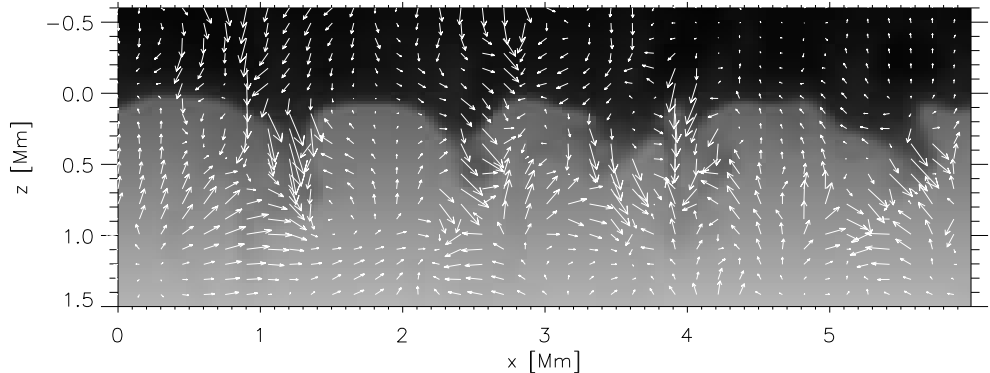}
\caption{Vertical slice from a 3D solar model atmosphere with
the velocity flow field indicated by arrows. The broad warm upflows
are characterized by rather slow motion while the cool material
is descending relatively rapidly in finger-like structures.
Note that the original simulation extends both above and below
the region shown here.}
\end{figure}

\subsection{Stellar Surface Convection
\label{simulations}}

A recent realisation which has emerged from convection simulations
like those described here is the fundamental importance of
radiative cooling in a thin surface layer for driving the convection
(Stein \& Nordlund 1998). As the ascending gas approaches the
optical surface radiation starts to escape, lowering the temperature
of the gas. In these regions the opacity is very sensitive to
the temperature, which means that the opacity decreases rapidly.
As a consequence more photons can leak out bringing the temperature
down further. 
This strong feedback causes a sudden drop in the gas temperature,
bringing the gas from around $10\,000$\,K to around $5\,000$\,K 
in about a minute for the Sun. 
The cooled, entropy-deficient material now has a negative buoyancy
and is therefore being accelerated downward. It is noteworthy that
almost all buoyancy work is done by the downflows rather than
the upflows, which instead are mainly being pushed upwards 
by mass conservation.
This picture is clearly conceptually very different from
the traditional view of convection as rising bubbles,
as in the mixing length theory or relatives thereof.

Granulation in other solar-type stars is qualitatively similar to
the solar case with broad upflows and concentrated downdrafts
according to numerical simulations.
This is also the case at low metallicities, but there are 
distinct differences in the optically thin layers.
The upflowing gas overshoots into the convectively stable
layers above the optical surface. The gas temperature is
then determined by a competition between radiative heating
and adiabatic cooling due to the expansion following the
density stratification. 
At solar metallicity, the large number of spectral lines (continuum
absorption is very inefficient in these optically thin layers)
retain the temperature close to the radiative equilibrium value. 
At low metallicity however, the severe lack of lines shifts
the balance towards adiabatic cooling, leading to distinct
sub-radiative equilibrium temperatures (Asplund et al. 1999).
The temperature difference between a 3D model and a corresponding
1D model can easily reach $1\,000$\,K in the upper layers,
which naturally have dramatic consequences for any spectral
lines formed in those regions, such as molecular lines.

\section{3D LTE Spectral Line Formation}

\subsection{Radiative Transfer}

Using the 3D convection simulations as a time-series of
3D model atmospheres, spectral line formation calculations
can be performed which subsequently can
be spatially and temporally averaged to produce disk-integrated
line profiles.
In LTE, the solution of the radiative transfer equation
is straightforward as the level populations are directly
obtained from the Boltzmann and Saha distributions.
This formal solution is computed using a Feautrier (1964) scheme
with long characteristics without any continuum scattering terms
but accounting for the Doppler shifts caused by the convective
motions in the simulations.
Before the radiative transfer calculations, the vertical resolution
is improved but the horizontal
resolution of the original simulation is decreased to typically
$N_{\rm x}*N_{\rm y} = 50*50$ to ease the computational burden,
which however does not encumber the results, as extensive tests show.
The calculation of a 3D flux profile require
some 100 wavelength points ($N_\lambda$), 10-20 angles ($N_{\rm angles}$) 
and at least 50 snapshots ($N_{\rm t}$)
to obtain statistically significant profiles in terms of line
asymmetries; for most abundance purposes significantly fewer snapshots
are necessary.
Thus a single 3D LTE flux profile require typically
$N_{\rm t}*N_{\rm x}*N_{\rm y}*N_{\rm angles}*N_\lambda \ga 10^8$
1D radiative transfer calculations, which however is still
achievable on modern workstations.
In addition,
the calculations are normally performed for three different element abundances
to enable interpolation to the requested line strength.
For computational reasons, the monochromatic continuous opacities
and equation-of-state data are pretabulated and interpolated during
the line calculations.
For each snapshot the spatially resolved
intensity profiles  (Fig. 2) for the different angles are stored,
from which a disk-integration can be performed to account for
rotational broadening (Dravins \& Nordlund 1990) in the final
spatially and temporally averaged flux profile. It is important to emphasize that
no micro- or macroturbulence enter the calculations.

\begin{figure}
\label{FeI6219_xy}
\plotfiddle{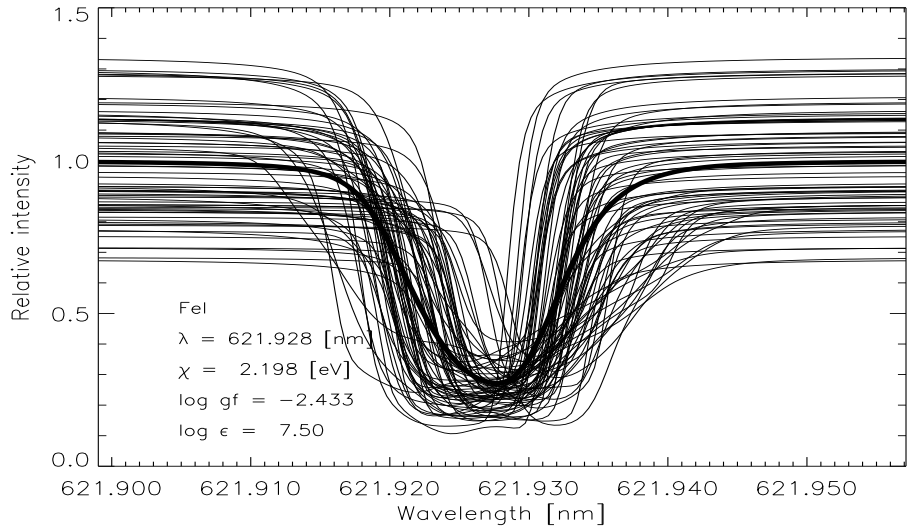}{7cm}{0}{75}{60}{-190}{0}
\caption{Spatially resolved profiles of the intermediate strong
Fe\,{\sc i} 621.9\,nm line
predicted in a 3D solar model atmosphere. Note that the core
may occasionally be affected by the artificial upper boundary. }
\end{figure}

\subsection{Spectral Line Shapes and Asymmetries}

\begin{figure}
\label{FeI6219_obs}
\plotfiddle{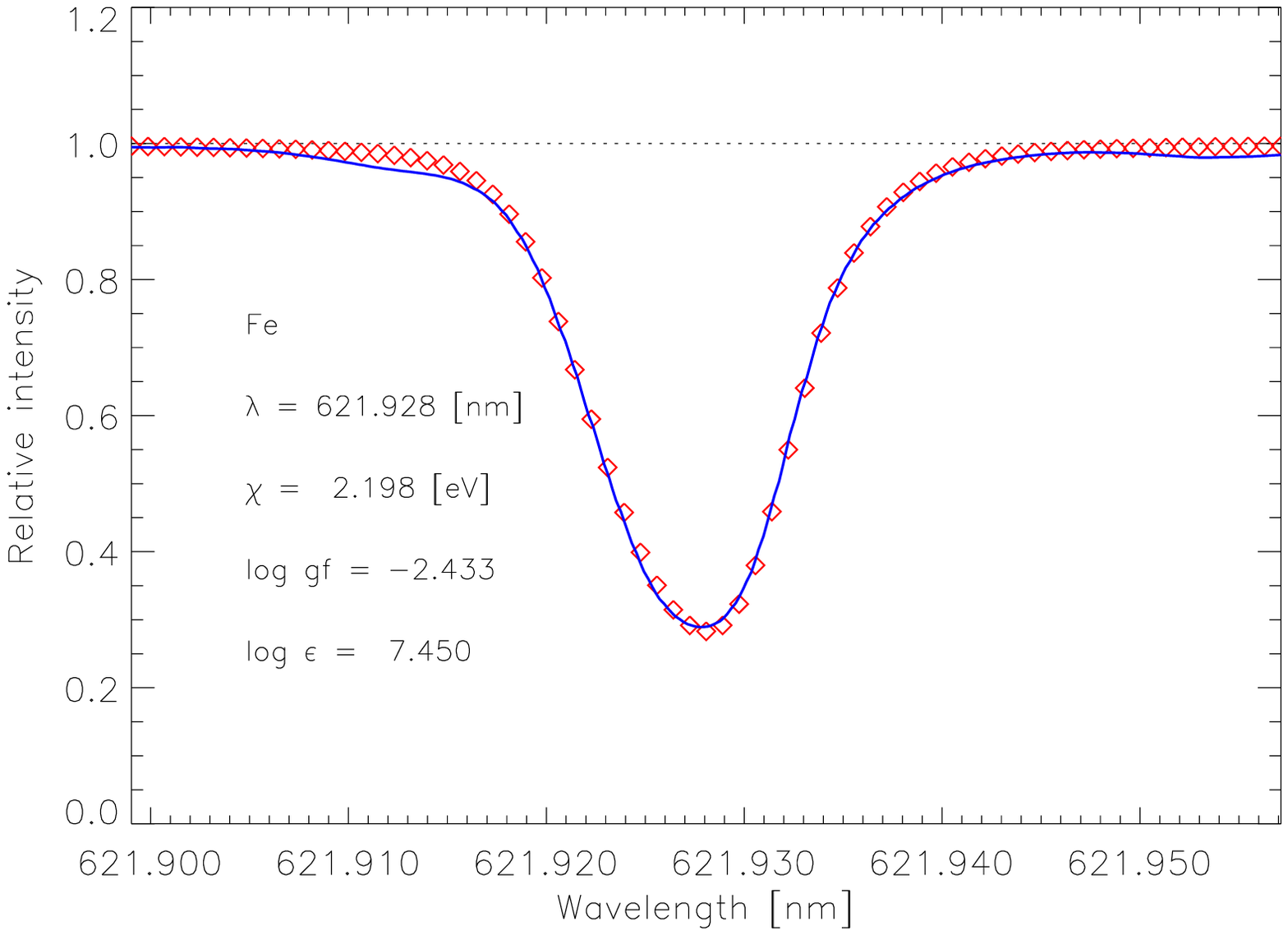}{8cm}{0}{75}{70}{-190}{0}
\plotfiddle{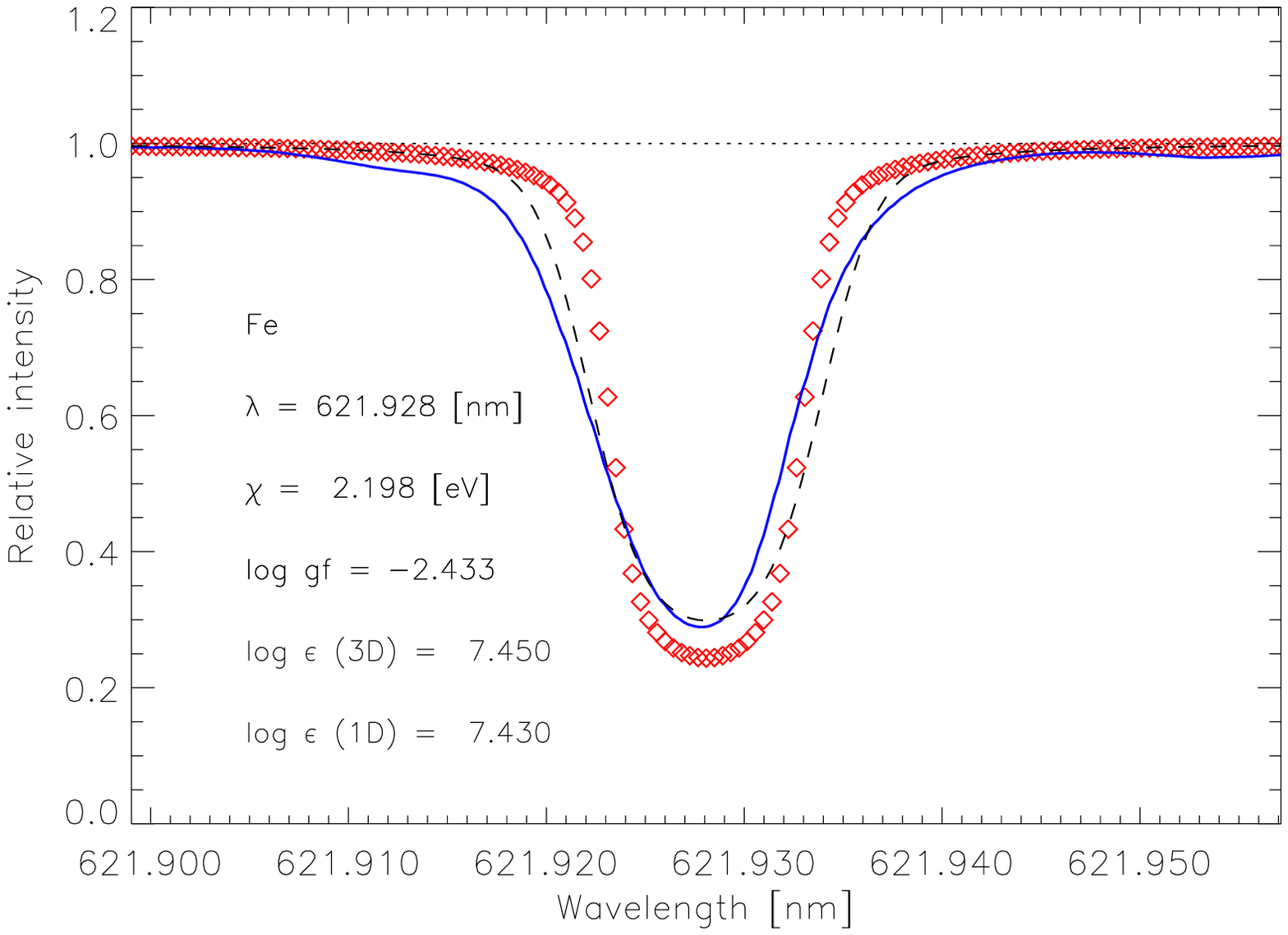}{8cm}{0}{75}{70}{-190}{0}
\caption{{\em Upper panel:} Predicted spatially and temporally averaged disk-center
intensity profile of
the Fe\,{\sc i} 621.9\,nm line in the Sun (diamonds) compared with 
observations (solid line). Note the slight problem in the core of this
saturated line and the blends in the blue and (far) red wings.
{\em Lower panel:} Same as above but when artificially removing all Doppler shifts
arising from the convective motions (diamonds). Also shown is an optimized
1D profile with micro- and (Gaussian) macroturbulence (dashed line)
producing the same line strength as the 3D profile in the upper panel.}
\end{figure}

The 3D hydrodynamical models described above may be more sophisticated than
existing 1D hydrostatic models but they are not necessarily more realistic.
Special efforts have therefore been made in comparing in great detail the
predictions with observations in terms of for example line shapes and asymmetries
to verify how reliable the simulations are.
Considering the amazing assortment of strengths, shapes and shifts of spatially
resolved profiles, it is obvious that the final averaged profile will be
an excellent probe of the atmospheric structure and a very challenging test
of the models.
Here, the Sun is a perfect test-bench as comparisons can be made without the interference
of rotational broadening and information is available for spatially resolved
profiles.

\begin{figure}[t!]
\label{Fe_bis}
\plotfiddle{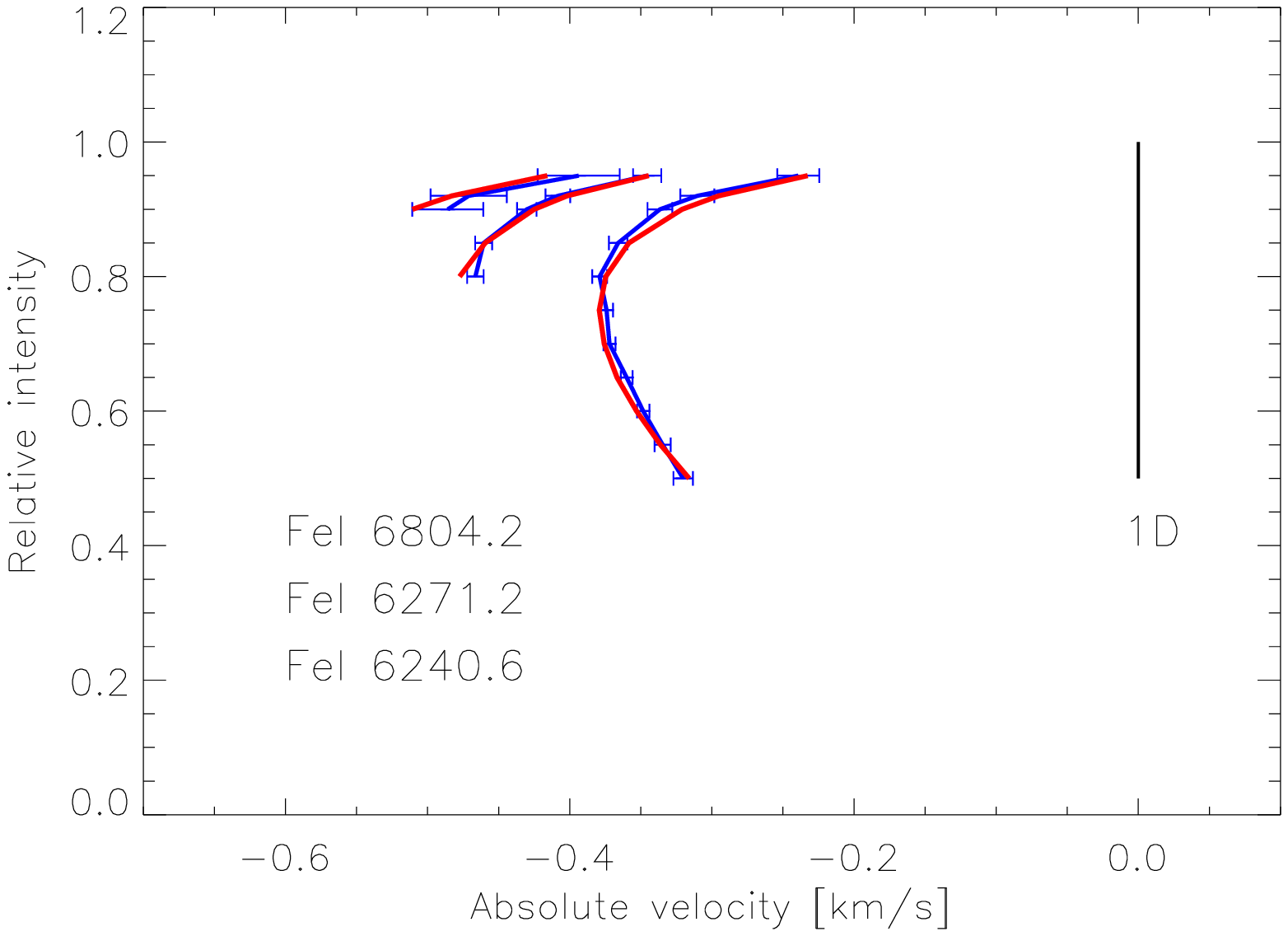}{7cm}{0}{75}{60}{-190}{00}
\caption{Examples of predicted (solid lines) and observed (solid lines with
error bars) spectral line bisectors of a few
Fe\,{\sc i} lines in the Sun. The agreement even on an absolute
wavelength scale is clearly highly satisfactory, while of course
the corresponding 1D profiles are purely symmetric. }
\end{figure}

As has been well-known since long, 1D models can neither predict the
strengths and shapes of lines without invoking the ad-hoc fudge factors
micro- and macroturbulence and of course fail utterly to describe
line asymmetries. In stark contrast, the 3D line profiles agree almost
perfectly with observations, as shown in Figs. 3 and 4 (see also Asplund et al. 2000b,
Nissen et al. 2000; Allende Prieto et al. 2002).
The self-consistently calculated Doppler shifts arising from the convective
motions and to a minor part the oscillatory motions
broaden the lines sufficiently. This is also evident from tests when artificially
removing all velocity information in the simulations (Fig. 3), giving profiles
which closely resemble 1D profiles without micro- and macroturbulence.
Thus, the concepts of micro- and macroturbulence
introduced in 1D analyses have nothing to with neither turbulence nor microscopic motions
but stems from gradients in the convective velocities which are resolved at the
present generation of 3D model atmospheres (Asplund et al. 2000a).
This is partly due to the fact that
the line formation process is heavily biased towards upflows
(large area coverage, high continuum intensity, steep temperature
gradients) where the divergent nature of the flow prevents
large amount of turbulence to develop in spite of the
high Reynolds-numbers of the plasma (Nordlund et al. 1997).
The excellent overall agreement between predicted and observed line profiles
in general makes it very easy to detect minor blends which otherwise may
well have gone unnoticed (Allende Prieto et al. 2001).

Even the detailed asymmetries and shifts of lines are very well reproduced
with the 3D simulations, both for the Sun and for other
solar-type stars (Asplund et al. 2000b; Allende Prieto et al. 2002),
as illustrated in Fig. 4.
The accuracy is typically $0.05-0.1$\,km\,s$^{-1}$ for weak lines, which is
similar to the uncertainties in the atomic data.
In fact, in a few cases the comparisons between predicted and observed
solar line profiles have revealed minor typographical errors in
tables of laboratory wavelengths.
For a large sample of weak solar Fe\,{\sc i} lines, Asplund et al. (2000b) found
an average difference of $0.000\pm 0.053$\,km\,s$^{-1}$ for the
observed and theoretical line shifts (with the gravitational redshift removed).
Strong lines show a systematic discrepancy in the line core which is
likely attributable to effects of the artificial upper boundary or a less
realistic temperature structure in the uppermost atmospheric layers.
Also the observed behaviour of spatially resolved solar line profiles
are well matched by the predictions from the simulations
(Kiselman \& Asplund, in preparation).



\section{3D Non-LTE Spectral Line Formation}

\subsection{Radiative Transfer}

Line formation does not of course obey LTE in general.
To make optimal use of the 3D hydrodynamical model atmospheres
described above it is necessary to develop tools that
can also handle detailed 3D non-LTE line formation.
This challenging task has only recently been
addressed with a very promising outlook for the future
(Kiselman 1997;
Uitenbroek 1998; Botnen \& Carlsson 1999; Asplund et al. 2002).
The major complication in any non-LTE calculation is of course that
the level populations depend on the radiation field which
in turn depends on the level populations.
The radiative transfer equation must therefore be solved
for all relevant wavelengths together with the rate
equations of the various levels, normally under the
simplifying assumption of time-independence (so-called statistical
equilibrium, ${\rm d}n_{\rm i}/{\rm d}t=0$).
To obtain a consistent result, an iterative procedure is adopted
using normally an accelerated lambda-iteration technique allowing convergence
typically within 10 iterations ($N_{\rm iter}$).
For a normal size model atom there may 100 radiative transitions ($N_{\rm tran}$),
each described with 50-100 wavelengths ($N_\lambda$).
In total, each 3D non-LTE calculations therefore require
$N_{\rm t}*N_{\rm x}*N_{\rm y}*N_{\rm angles}*N_{\rm tran}*N_\lambda*N_{\rm iter}$
corresponding  1D radiative transfer solutions.
Hence it should come as no surprise that the formal solution
of the radiative transfer equation with the current estimate of the
source function is by far the most dominating task in the calculations.
It is not possible to perform such 3D non-LTE calculations for long
time-series as can be done in LTE and instead the computations are restricted
to a few individual snapshots. Fortunately, 3D non-LTE effects are quite robust
entities which differ little from snapshot to snapshot, as verified by
various test calculations.

The 3D non-LTE code {\sc multi3d} (Botnen \& Carlsson 1999; Asplund et al. 2002)
is essentially a 3D version of the widely used {\sc multi}-code for
1D statistical equilibrium calculations (Carlsson 1986).
The formal solution of the radiative transfer equation is computed
using a short-characteristic method utilizing the horizontal boundary
conditions of the 3D atmospheres and accounting for Doppler shifts introduced
by the convective motions.
The equation-of-state and continuous opacities are provided by the Uppsala
package (Gustafsson et al. 1975 and subsequent updates) and
allowance for line-blanketing in the photo-ionization rates is possible.

\begin{figure}[t!]
\label{LiI_IW}
\plotone{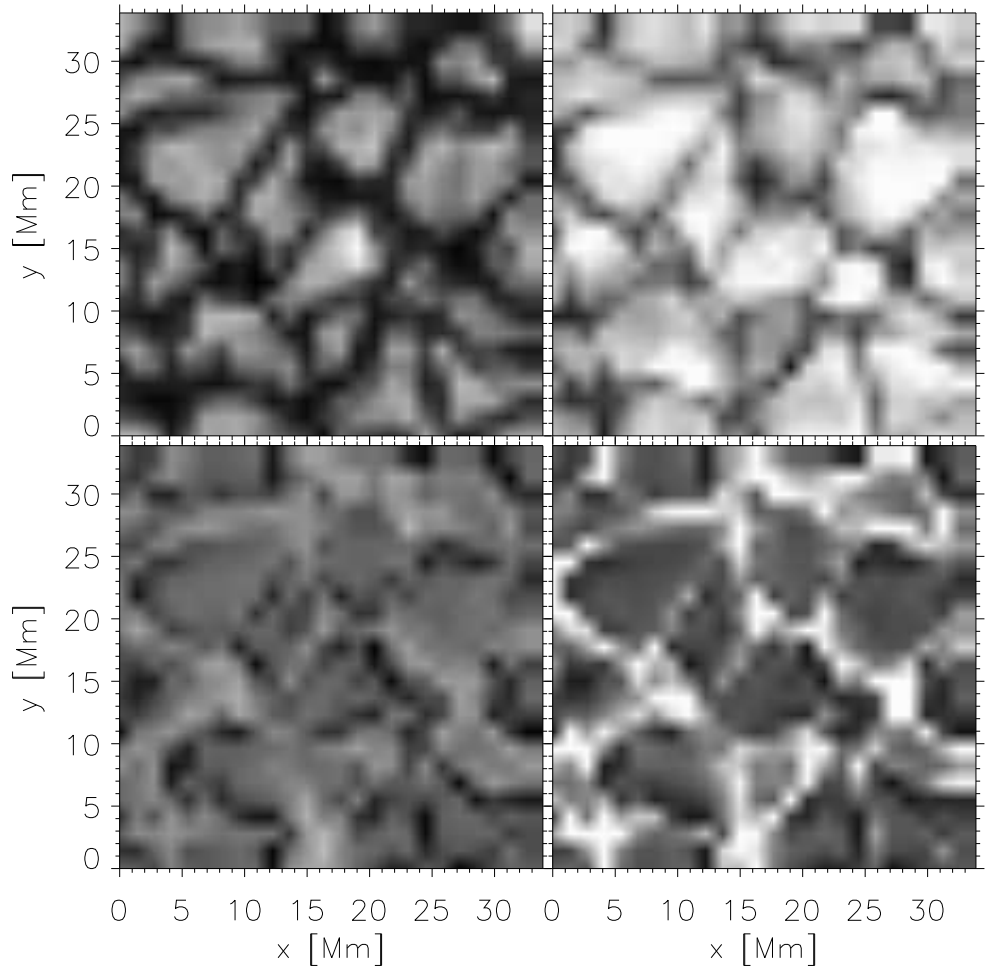}
\caption{Granulation pattern in HD\,140283
seen in disk-center continuum intensity ({\it upper left panel}) and
equivalent width of the Li\,{\sc i} 670.8\,nm line in
LTE ({\it upper right panel}) and non-LTE ({\it lower left panel}).
Also shown is the ratio of the non-LTE and LTE equivalent widths
({\it lower right panel}).}
\end{figure}


\subsection{Li Abundances in Metal-Poor Stars}

Given the possibility to estimate the total baryon density of the Universe
using the Li abundances in pristine metal-poor halo stars, a large
effort has gone into this endeavour ever since the first detection
of the Li\,{\sc i} 670.8\,nm resonance line in such stars
(Spite \& Spite 1982; Ryan et al. 1999).
In view of the large differences in the temperature structures of
3D and 1D models of halo stars (Asplund et al. 1999, see also Sect. \ref{simulations}),
there is clearly
a great need to study the Li line formation in detail in 3D model
atmospheres.
Indeed, preliminary calculations assuming LTE indicated that existing
1D analyses had overestimated the primordial Li abundance by as much
as $0.2-0.35$\,dex (Asplund et al. 1999). Extreme caution must however
be exercised in interpreting these results as Li
may be prone to severe over-ionization effects which
could compensate the 3D LTE corrections. Indeed our new 3D non-LTE
calculations reveal pronounced 3D non-LTE effects which bring the
derived Li abundances almost exactly back to those determined with 1D models
(Asplund et al. 2002).

\begin{figure}[t!]
\label{LiI_WvsI}
\plotfiddle{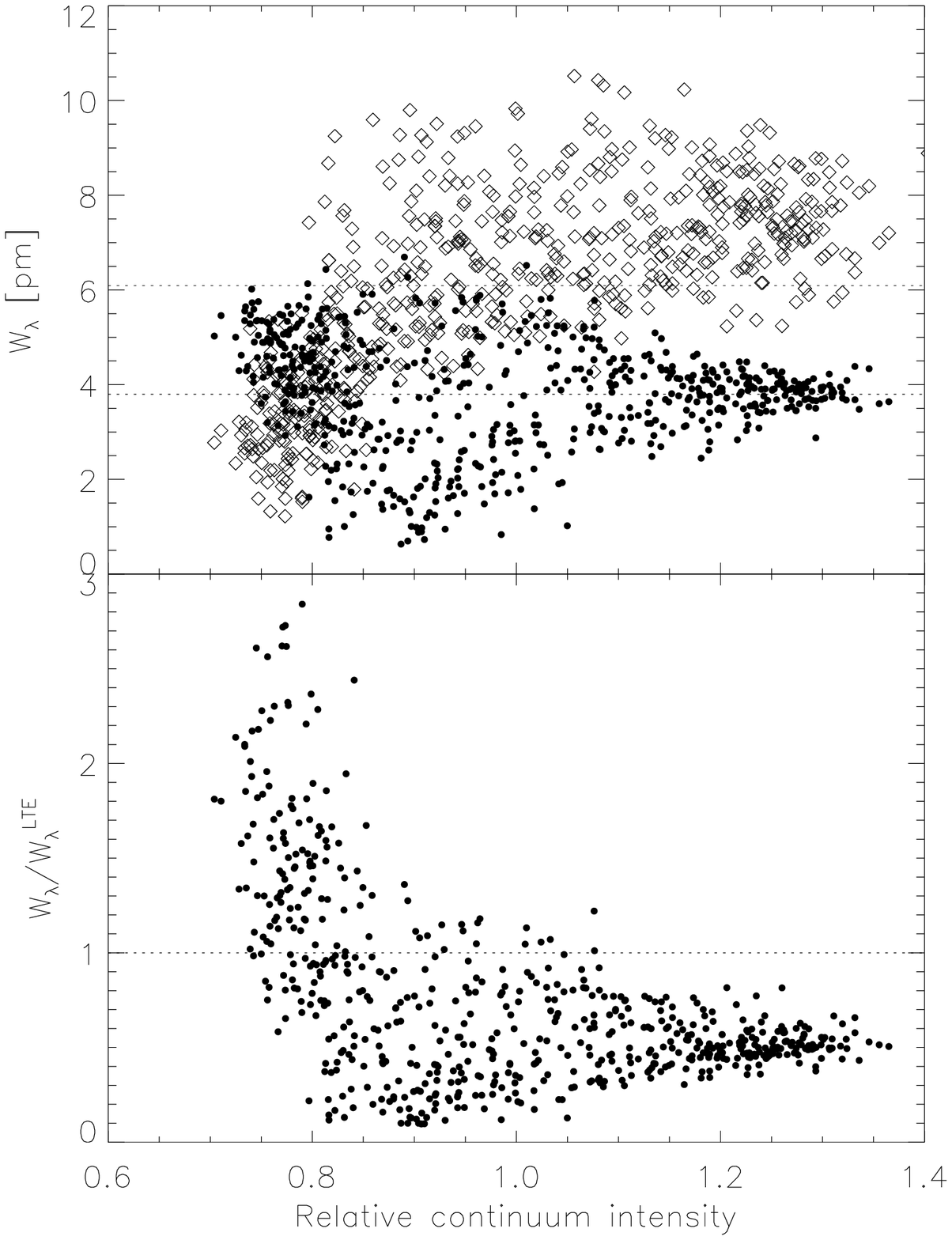}{14cm}{0}{80}{70}{-210}{-20}
\caption{{\em Upper panel:} 
Predicted disk-center intensity equivalent widths
in HD\,140283 in non-LTE (filled circles) and LTE (open diamonds)
across the granulation pattern.
{\em Lower panel:} Ratio of non-LTE to LTE equivalent widths.}
\end{figure}

With {\sc multi3d} we have recently investigated possible departures from
LTE for Li in the two halo stars HD\,140283 and HD\,84937 using a 21-level
Li model atom with 90 radiative transitions (Carlsson et al. 1994).
Colllisional excitation and ionization due to electrons are accounted for
but not the corresponding case of hydrogen collisions, which however have
limited impact on the final results.
Compared with the LTE case, the Li line is much weaker in non-LTE,
as shown in Figs. 5 and 6. In LTE the line is generally strong above the warm upflows
while this is not the case in non-LTE, which implies that the population density
is largely controlled by the radiation field rather than the local temperature.
As anticipated, the main non-LTE mechanism is over-ionization, which
decreases the line opacity significantly.
In terms of abundances, the 3D non-LTE calculations yield Li abundances
well within 0.05\,dex to the standard 1D LTE case, at least for these
two halo stars.
It is noteworthy though that the subordinate 610.4\,nm line show an
abundance correction of about $+0.15$\,dex compared with 1D, which should
be detectable with high enough $S/N$ observations.
It would be premature however from these limited calculations to conclude
that the apparent trend in Li abundances with metallicity derived with
1D models (Ryan et al. 1999) is due to galactic chemical evolution of Li.
The exact magnitude of the 3D non-LTE effect will depend on the
stellar parameters, in particular $T_{\rm eff}$ and [Fe/H].
We are planning to investigate the metallicity dependence of the 3D corrections
in the near future.

\section{Concluding Remarks}

With realistic 3D hydrodynamical model atmospheres of late-type stars and
corresponding 3D LTE and non-LTE spectral line formation tools now becoming
available many potential systematic errors in standard analyses can finally be removed
eliminated.
In many cases the results may well be dramatically revised
(e.g. Asplund \& Garc\'{\i}a P{\'e}rez 2001; Allende Prieto et al. 2001; Nissen et al. 2002),
which can have large impact on the inferred conclusions on stellar, galactic
and cosmic evolution. However, the significant efforts in pursuing this
ambitious endeavour are justified also for those cases where 3D
only confirms previous 1D results as the uncertainties are greatly reduced.
The reliability of existing results can clearly only be judged
after more sophisticated calculations have been performed.

\acknowledgments

We are greatly indebted to the continuing efforts of a large
number of wonderful collaborators in the quest of developing
3D model atmospheres and 3D line formation and their applications to
astrophysical problems.
In particular we would like to mention Carlos Allende Prieto, Mats Carlsson,
Ana Garc\'{\i}a P{\'e}rez, Nicolas Grevesse, Dan Kiselman,
David Lambert, Poul Erik Nissen, \AA ke Nordlund,
Francesca Primas, Jacques Sauval, Bob Stein, and Regner Trampedach.
Finally, we are grateful to the editors for their patience.

\end{document}